\documentclass[twocolumn,showpacs,preprintnumbers,amsmath,amssymb]{revtex4}

\usepackage{graphicx,subfigure}
\usepackage{dcolumn}
\usepackage{bm}

\begin{document}

\preprint{APS/123-QED}

\title{Is the Scaling of Supersonic Turbulence Universal?}

\author{Wolfram Schmidt}
\email{schmidt@astro.uni-wuerzburg.de}
\affiliation{Lehrstuhl f\"{u}r Astronomie, Institut f\"{u}r
Theoretische Physik und Astrophysik, Universit\"{a}t W\"{u}rzburg,
Am Hubland, D-97074 W\"{u}rzburg, Germany} 

\author{Christoph Federrath}
\email{chfeder@ita.uni-heidelberg.de}
\author{Ralf Klessen}
\affiliation{Institut
f\"{u}r Theoretische Astrophysik, Universit\"{a}t Heidelberg,
Albert-Ueberle-Str.~2, D-69120 Heidelberg, Germany}

\date{\today}

\begin{abstract}

The statistical properties of turbulence are considered to be
universal at sufficiently small length scales, i.~e., independent of
boundary conditions and large-scale forces acting on the
fluid. Analyzing data from numerical simulations of supersonic
turbulent flow driven by external forcing, we demonstrate that this is
not generally true for the two-point velocity statistics of
compressible turbulence. However, a reformulation of the refined
similarity hypothesis in terms of the mass-weighted velocity
$\rho^{1/3}\bm{v}$ yields scaling laws that are almost insensitive to
the forcing. The results imply that the most intermittent
dissipative structures are shocks closely following the scaling of
 Burgers turbulence.

\end{abstract}

\maketitle

The notion of universality is central to theoretical and observational
accounts of turbulence. For incompressible turbulence, the
mathematical analysis carried out by Kolmogorov led to the famous
$2/3$ law for the second order structure function of turbulent
velocity fluctuations \cite{Kolmog41,Frisch}. Remarkably, this scaling
law was experimentally confirmed even if the premises of Kolmogorov's
theory--statistical equilibrium, homogeneity and isotropy--were not
satisfied. It is commonly accepted that this is due to an inertial
subrange of scales, where the dynamics of turbulence conditions the
flow such that these premises are met asymptotically toward smaller
length scales independent of the large-scale properties of the
flow. Whereas terrestrial applications are mostly concerned with the
incompressible regime, the idea that universal scaling also exists in
the inertial subrange of the highly compressible, supersonic
turbulence has become popular in astrophysics. Particularly,
observational properties of star-forming clouds are explained
by supersonic turbulent motion which is seen as an agent that controls
the formation of stars besides gravity \cite{LowKless04}. We 
tested the hypothesis of universality on data from numerical
simulations of supersonic turbulence with the same root mean square
(RMS) Mach numbers but different large-scale forcing. The result is
that universality in the sense of the Kolmogorov theory is clearly
violated even if intermittency corrections are applied, but scaling
laws that are nearly indepenent of the forcing apply to the mass-weighted velocity
$\tilde{\bm{v}}=\rho^{1/3}\bm{v}$ introduced by Kritsuk et
al. \cite{KritPad07}. Moreover, the scaling exponents turn out to be
consistent with log-Poisson models, if the \emph{most intermittent}
dissipative structures are shocks fulfilling the scaling of Burgers
turbulence.

The fundamental relation for the scaling properties of incompressible turbulence is the refined similarity hypothesis \cite{Kolmog62,Frisch},
\begin{equation}
	\label{eq:ref_sim}
	S_{p}(\ell)=C_{p}\ell^{p/3}\langle\epsilon_{\ell}^{p/3}\rangle,
\end{equation}
where the $C_{p}$ are constant dimensionless coefficients, $S_{p}(\ell):=\langle\delta v_{\ell}^{p}\rangle$ are the statistical moments of the velocity fluctuation $\delta v_{\ell}=|\bm{v}(\bm{x},t)-\bm{v}(\bm{x}+\bm{\ell},t)|$ of order $p=1,2,3,\ldots$ and $\epsilon_{\ell}$ is the rate of energy dissipation per unit mass averaged over a region of size $\ell$. The brackets $\langle\ \rangle$ denote the ensemble average. The factor $\langle\epsilon^{p/3}_{\ell}\rangle\propto\ell^{\tau_{p/3}}$ is attributed to the intermittency of turbulence \cite{Frisch}. The
scaling exponents of $S_{p}(\ell)$ in the inertial subrange are thus given by $\zeta_{p}=p/3+\tau_{p/3}$. 

Modeling turbulent energy dissipation by a random cascade obeying log-Poisson statistics,
Dubrulle \cite{Dub94} showed that the relative scaling exponents $Z_{p}$ have the general
form
\begin{equation}
	\label{eq:scl_expn}
	Z_{p}:=\frac{\zeta_{p}}{\zeta_{3}}=(1-\Delta)\frac{p}{3}+\frac{\Delta}{1-\beta}\left(1 - \beta^{p/3}\right) .	
\end{equation}
The intermittency parameter $\beta$ is interpreted as a random cascade factor relating dissipative
structures of different intensity, and $C=\Delta/(1-\beta)$ is the co-dimension of the most intense dissipative structures \cite{SheWay95}. The co-dimension is related to the fractal dimension by
$C=D-3$. As argued by She and L\'{e}v\^{e}que \cite{SheLev94}, the scaling of these structures, $\ell^{-\Delta}$, is given by the inverse of the kinetic energy available for dissipation at the length scale $\ell$. For incompressible turbulence, the most intense dissipative structures are assumed to be vortex filaments, for which $C=2$ and $\Delta=2/3$.

Boldyrev \cite{Boldyrev02} proposed $C=1$ for supersonic turbulence (Kolmogorov-Burgers model), because he considered the most intense dissipative structures to be shocks, while keeping $\Delta=2/3$ as in the She-L\'{e}v\^{e}que model for incompressible turbulence. However, the kinetic energy at the length scale $\ell$ is proportional to $\ell$ for Burgers turbulence ($\delta v_{\ell}\propto \ell^{1/2}$).
Following the arguments by She and L\'{e}v\^{e}que we propose that the
most intense dissipative structures should obey the scaling law $\ell^{-1}$ rather than $\ell^{-2/3}$, 
i.~e., $\Delta = 1$. The scaling exponents obtained from equation~(\ref{eq:scl_expn}) for $\Delta=1$ are markedly different from the prediction of the Kolmogorov-Burgers model. 

In the following, we will determine the relative scaling exponents $Z_{p}:=
\zeta_{p}/\zeta_{3}$ from the relations $S_{p}(\ell)=S_{3}(\ell)\ell^{Z_{p}}$ \cite{BenzCil93}
for simulations of supersonic isothermal turbulence with periodic large-scale stochastic forcing
\cite{FederSchm08}. In these simulations, the compressible Euler
equations were solved with the piecewise parabolic method in the
FLASH3 implementation \cite{ColWood84,FryOls00}. In one case, purely
solenoidal (divergence-free) stochastic forcing was applied, in the other case
the forcing was purely compressive (rotation-free)
\cite{EswaPope88,SchmHille06}. We use the term compressive synonymous
to dilatational (rotation-free). In each case, the system was evolved
over 10 auto-correlation time scales (subsequently denoted by $T$) of
the force field at grid resolutions $N=256^{3}$, $512^{3}$ and
$1024^{3}$. Turbulence was found to be fully developed with a
steady-state RMS Mach number $\approx 5.5$ after about two
autocorrelation time scales. Following previous numerical studies of
supersonic turbulence \cite{PadJim04,KritNor07}, we
computed transversal structure functions $S_{p}^{\perp}(\ell)$, i.~e.,
the $p$-th moments of the velocity fluctuation projected perpendicular
to the spatial separation $\bm{\ell}$. The structure functions were
computed from a statistically converged sample in the interval $2\le t/T\le 10$ using a
Monte Carlo algorithm  \cite{FederSchm08}.

\begin{figure}
    \includegraphics[width=\linewidth]{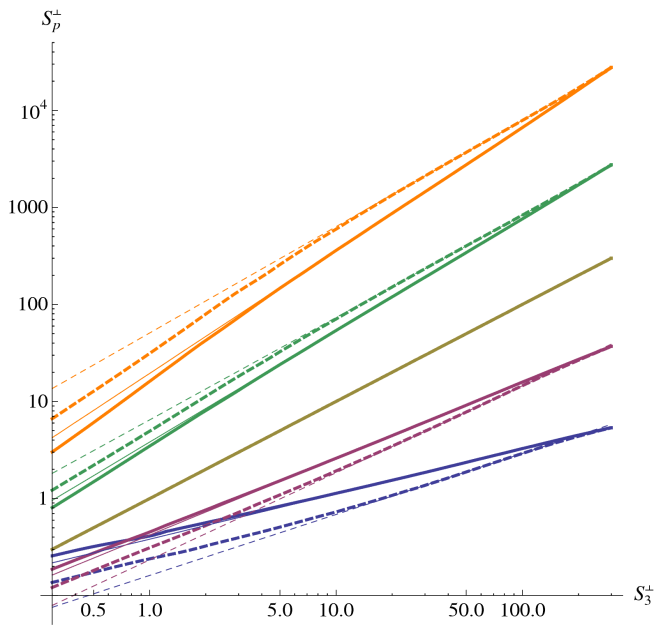}
    \caption{\label{fig:StransESS} Interpolated relations between the time-averaged 
    structure functions $S_{p}^{\perp}$ and $S_{3}^{\perp}$ for
    solenoidal (solid lines) and compressive (dashed lines)
    forcing. The order $p$ ranges form $1$ (bottom) to $5$ (top).
    The corresponding power-law fit functions are plotted as thin lines.}
\end{figure}

The structure functions $S_{p}^{\perp}(\ell)$ averaged over the time interval $2\le t/T\le
10$ for $N=1024^{3}$ are plotted as functions of the time-averaged third-order structure
function $S_{3}^{\perp}(\ell)$ in Fig.~\ref{fig:StransESS}.
The exponents $Z_{p}$ are given by the slope 
of $\log S_{p}^{\perp}$ vs. $\log S_{3}^{\perp}$. For the
determination of linear fit functions, $\mathrm{fit}_{p}\propto
Z_{p}\log S_{3}^{\perp}$, we imposed the criterion $\forall p\le
5:\mathrm{err}_{p}:=|\exp{(\mathrm{fit}_{p})}-S_{p}^{\perp}|/S_{p}^{\perp}<
0.01$ in the fit range. This error criterion is fulfilled in the intervals 
$12.0\le S_{3}^{\perp}\le 120$ for
solenoidal forcing and $25.0\le S_{3}^{\perp}\le 150$ for compressive
forcing. One should note that the relations
between $S_{p}^{\perp}(\ell)$ and $S_{3}^{\perp}(\ell)$ agree quite
closely with the fit functions  ($\mathrm{err}_{p}<0.05$) for large length scales
corresponding to the maximal values of $S_{3}^{\perp}$.

\begin{figure*}
    \mbox{\subfigure[\ solenoidal]{\includegraphics[width=0.5\linewidth]{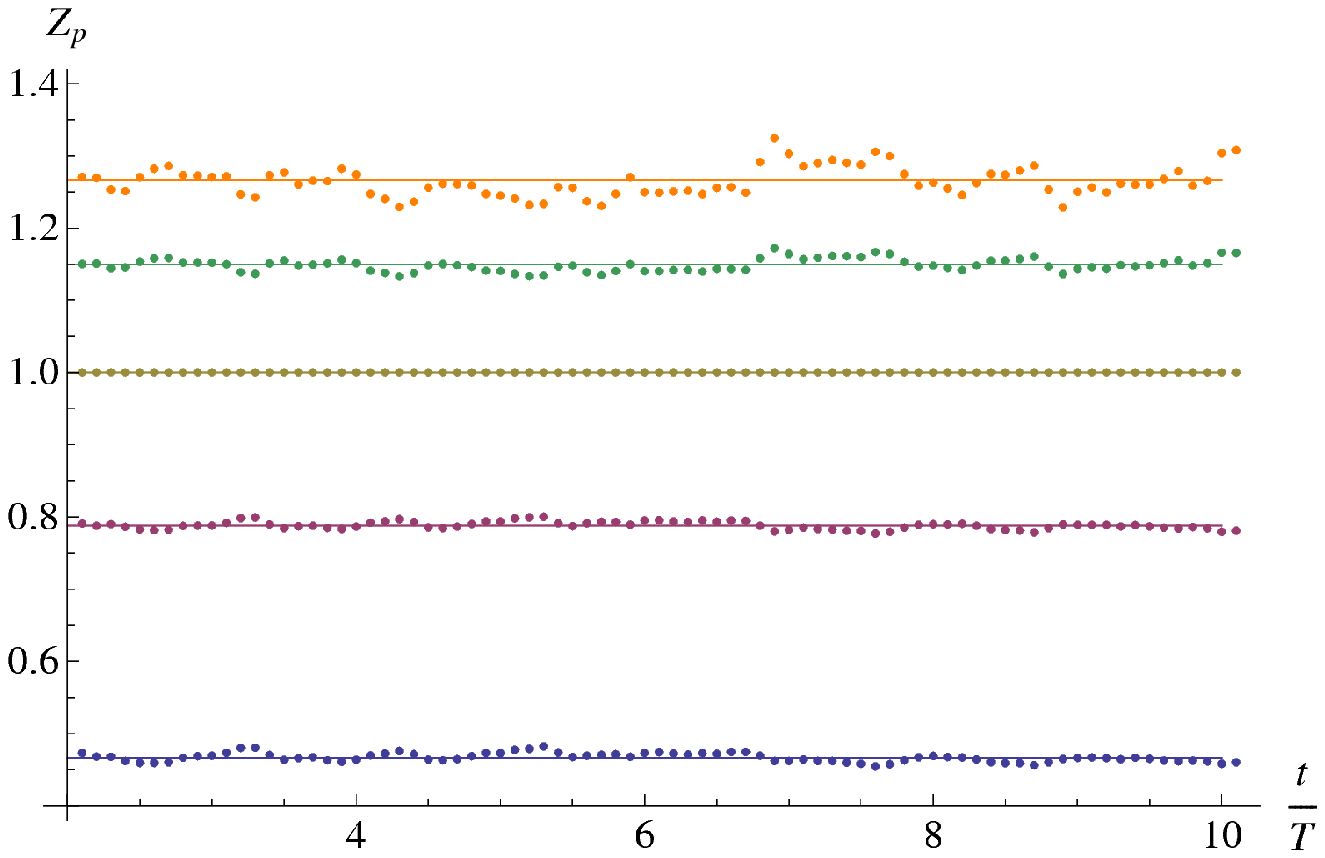}}\quad
                \subfigure[\ compressive]{\includegraphics[width=0.5\linewidth]{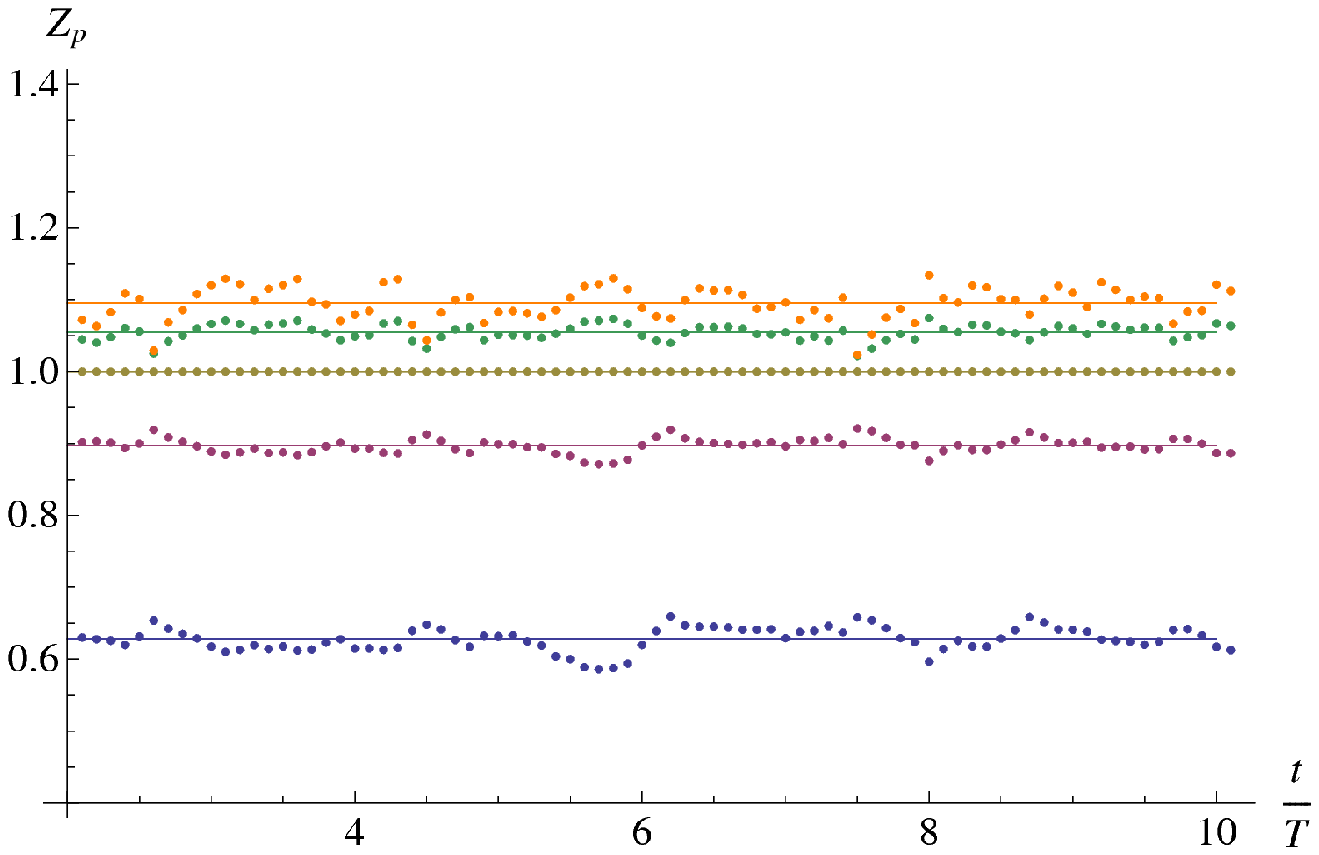}} }
\caption{\label{fig:Z} Time evolution of the relative scaling exponents $Z_{p}$. The time
	averages are indicated by the horizontal lines. }
\end{figure*}

The relative scaling exponents $Z_{p}$ inferred from the time-averaged
transversal structure functions are listed in Table~\ref{tab:Z}. 
In the case $N=1024^{3}$, the standard errors of the
parameters $Z_{p}$ are of the order $10^{-3}$. We estimate systematic
errors to be of the order $10^{-2}$. In Fig.~\ref{fig:Z}, we compare these
values to the instantaneous scaling exponents as functions of time for
$N=1024^{3}$. Most importantly, we see that the scaling exponents
resulting from solenoidal forcing differ markedly from the case of
compressive forcing. In each case, the variation of $Z_{1}$ and
$Z_{2}$ over the time scale $T$ clearly shows temporal correlation and there
appears to be anti-correlation with the scaling exponents of order
greater than three. This suggests that the instantaneous scaling
exponents show an imprint of the stochastic variation of the large
scale forcing rather than purely statistical scatter. Even for incompressible turbulence, 
an influence of the large scales on much smaller scales is reported \cite{AlexaMin05}.

\begin{table}
\caption{\label{tab:Z} Relative scaling exponents $Z_{p}$ from fits of time-averaged
	structure functions $S_{p}^{\perp}$ vs. $S_{3}^{\perp}$.}
\begin{ruledtabular}
\begin{tabular}{ccccccc}
 & $N$ & $Z_{1}$ & $Z_{2}$ & $Z_{3}$ & $Z_{4}$ & $Z_{5}$  \\
\hline
     &  $256^3$ & 0.472 & 0.786 & 1. & 1.160 & 1.289 \\
sol. &  $512^3$ & 0.474 & 0.792 & 1. & 1.149 & 1.265 \\
     & $1024^3$ & 0.466 & 0.788 & 1. & 1.150 & 1.266 \\
\hline
      &  $256^3$ & 0.603 & 0.879 & 1. & 1.072 & 1.126 \\
comp. &  $512^3$ & 0.627 & 0.896 & 1. & 1.056 & 1.097 \\
      & $1024^3$ & 0.628 & 0.897 & 1. & 1.055 & 1.095 \\
\end{tabular}
\end{ruledtabular}
\end{table}

\begin{figure}
    \includegraphics[width=\linewidth]{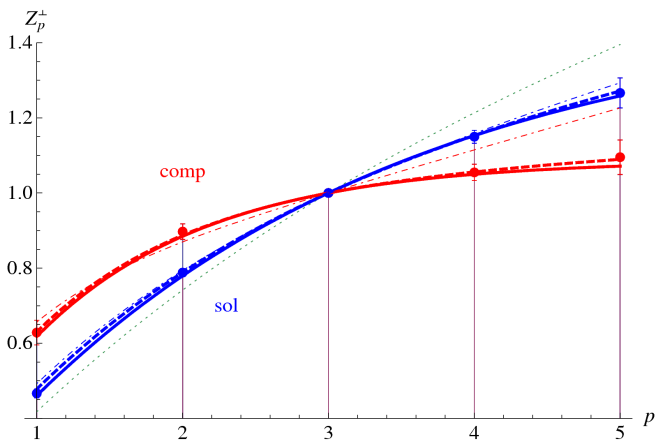}
    \caption{\label{fig:Z_model} Comparison of the relative scaling exponents $Z_{p}$ for solenoidal and compressive forcing (see Table~\ref{tab:Z}) with the Kolmogorov-Burgers model (dotted line) and log-Poisson model fits with $\Delta=2/3$ (dot-dashed lines), $C=1$ (thick dashed lines) and $\Delta=1$ (solid lines). The vertical bars indicate the standard deviations of the instantaneous values.}
\end{figure}

\begin{figure}
    \includegraphics[width=\linewidth]{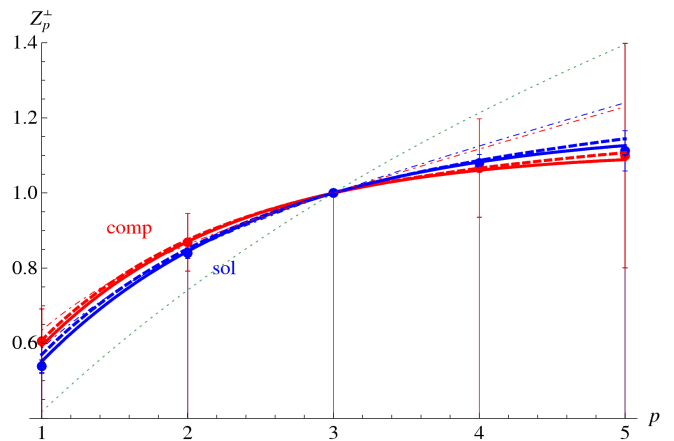}
    \caption{\label{fig:Z_model_rho3} Comparison of the relative scaling exponents
 $\tilde{Z}_{p}$ (mass-weighing $\tilde{\bm{v}}=\rho^{1/3}\bm{v}$) for solenoidal and compressive forcing (see Table~\ref{tab:Z_rho3}) with various models analogous to
    Fig.~\ref{fig:Z_model}.}
\end{figure}

The Kolmogorov-Burgers model implied by
equation~(\ref{eq:scl_expn}) for $\Delta=2/3$ and
$\beta=1-\Delta/C=1/3$ is plotted as dotted line together with our
data in Fig.~\ref{fig:Z_model}.  Clearly, there are large deviations
both for solenoidal and for compressive forcing. One reason is that
the Kolmogorov-Burgers model only applies in the hypersonic limit,
whereas the most intense dissipative structures are constituted by
varying fractions of vortex filaments and shocks depending on the RMS
Mach number. Then one would expect $1\le C\le 2$
\cite{PadJim04}. Moreover, fitting the log-Poisson model with
$\Delta=2/3$, we find $C_{\mathrm{sol}}\approx 0.76$ and
$C_{\mathrm{comp}}\approx 0.67$, which is about the minimal
co-dimension $C=\Delta=2/3$, for which the $Z_{p}$ are real. In the
case of compressive forcing, no closely matching fit function exists.
On the other hand, fitting the one-parameter family of models with
$\Delta=1$,  we obtain co-dimensions $C_{\mathrm{sol}}\approx 1.5$ and
$C_{\mathrm{comp}}\approx 1.1$. These models match the
time-averaged relative scaling exponents very well. We cannot fully
discriminate other families of models though. For instance, assuming 
$C=1$ as in the Kolmogorov-Burgers model and varying $\Delta$ as fit parameter, 
yields $\Delta_{\mathrm{sol}}=0.79$ and $\Delta_{\mathrm{comp}}\approx 0.94$. 
Nevertheless, in the case of compressive forcing, the match with the numerically computed scaling exponents is much better in comparison to the $\Delta=2/3$ models.

\begin{figure}
    \includegraphics[width=\linewidth]{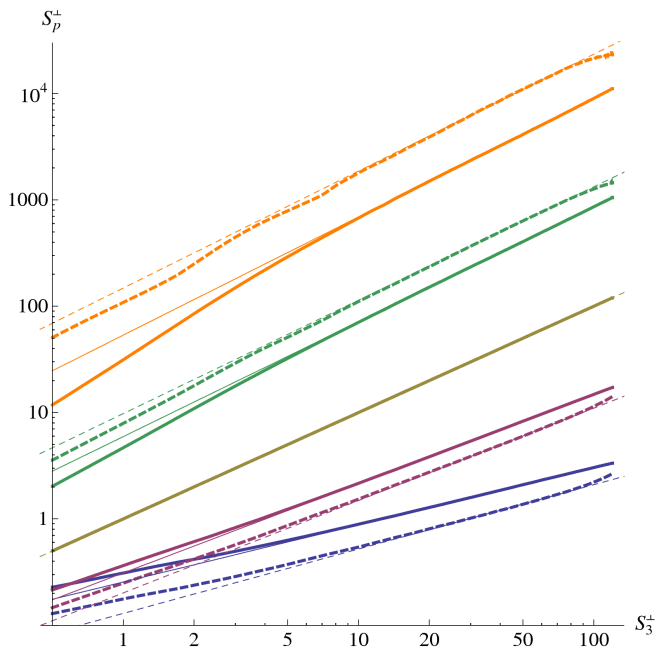}
    \caption{\label{fig:StransESS_rho3} Relations between the time-averaged 
    structure functions  $\tilde{S}_{p}^{\perp}$ and $\tilde{S}_{3}^{\perp}$ 
    with mass weighing analogous to Fig.~\ref{fig:StransESS}.}
\end{figure}

To carry over the log-Poisson models to compressible turbulence, the refined similarity hypothesis might
be applied in the form~(\ref{eq:ref_sim}), where the rate of energy dissipation at length scale $\ell$ is understood to be the Favre-filtered quantity $\epsilon_{\ell}=\langle\rho\epsilon\rangle_{\ell}/\rho_{\ell}$. The filter operation $\langle\ \rangle_{\ell}$ smoothes out fluctuations at length scales $\lesssim\ell$, $\rho_{\ell}:=\langle\rho\rangle_{\ell}$ and $\epsilon(\bm{x},t)$ is the local rate of energy dissipation per unit mass. On the other hand, one might consider the rate of dissipation per unit volume, $\tilde{\epsilon}_{\ell}:=\rho_{\ell}\epsilon_{\ell}$, as the variable from which the hierarchy should be constructed \cite{Fleck96}. This is also suggested by the alternative formulation of the log-Poisson model by She and Waymire \cite{SheWay95}. Following this proposition,
the refined similarity hypothesis with mass-weighing,
\begin{equation}
	\label{eq:ref_sim_mw}
	\tilde{S}_{p}(\ell)=\tilde{C}_{p}\ell^{\tilde{\zeta}_{3}p/3}\langle(\rho_{\ell}\epsilon_{\ell})^{p/3}\rangle^{\tilde{\zeta}_{3}},
\end{equation}
where $\tilde{S}_{p}(\ell):=\langle\delta (\rho^{1/3}v)_{\ell}^{p}\rangle$, leads
to the relative scalings $\tilde{S}_{p}(\ell)=\tilde{S}_{3}(\ell)\ell^{\tilde{Z}_{p}}$
for any $\tilde{\zeta}_{3}> 0$. 
As proposed by Kritsuk et al.\ \cite{KritPad07}, the exponents $\tilde{Z}_{p}$ associated
with the two-point statistics of $\tilde{\bm{v}}:=\rho^{1/3}\bm{v}$, are then given by an
expression analogous to~(\ref{eq:scl_expn}) with parameters $\tilde{\beta}$ and $\tilde{\Delta}$.

\begin{table}
\caption{\label{tab:Z_rho3} Relative scaling exponents $\tilde{Z}_{p}$ from fits of time-averaged
	mass-weighted structure functions $\tilde{S}_{p}^{\perp}$ vs. $\tilde{S}_{3}^{\perp}$
	with mass weighing $\tilde{\bm{v}}=\rho^{1/3}\bm{v}$.}
\begin{ruledtabular}
\begin{tabular}{ccccccc}
 & $N$ & $\tilde{Z}_{1}$ & $\tilde{Z}_{2}$ & $\tilde{Z}_{3}$ & $\tilde{Z}_{4}$ & $\tilde{Z}_{5}$\\
\hline
     &  $256^3$ & 0.546 & 0.839 & 1. & 1.094 & 1.150\\
sol. &  $512^3$ & 0.550 & 0.845 & 1. & 1.082 & 1.122\\
     & $1024^3$ & 0.539 & 0.840 & 1. & 1.080 & 1.112\\
\hline
      &  $256^3$ & 0.635 & 0.893 & 1. & 1.034 & 1.026\\
comp. &  $512^3$ & 0.634 & 0.887 & 1. & 1.050 & 1.068\\
      & $1024^3$ & 0.605 & 0.869 & 1. & 1.066 & 1.100\\
\end{tabular}
\end{ruledtabular}
\end{table}

The mass-weighted structure
functions $\tilde{S}_{p}^{\perp}(\ell)$
vs. $\tilde{S}_{3}^{\perp}(\ell)$ computed from our numerical data are plotted in Fig.~\ref{fig:StransESS_rho3}. In order to satisfy the chosen error
criterion, we had to restrict the fit functions to narrower ranges
$15.0\le\tilde{S}_{3}^{\perp}\le100$ and
$25.0\le\tilde{S}_{3}^{\perp}\le75.0$ for solenoidal and compressive
forcing, respectively. The resulting values of the relative scaling
exponents $\tilde{Z}_{p}$ are summarized in
Table~\ref{tab:Z_rho3}. Although the sensitivity on resolution is more
pronounced than for the $Z_{p}$ and the scatter of the instantaneous
values of $\tilde{Z}_{p}$ is very large due to the
pronounced intermittency of the mass density \cite{KritNor07},
trends are nevertheless discernable. Most noticeably, the differences between turbulence
driven by solenoidal and compressive forcing are substantially
reduced, especially, for the higher-order exponents.  As shown in
Fig.~\ref{fig:Z_model_rho3}, again neither the Kolmogorov-Burgers
model nor log-Poisson models with $\Delta=2/3$ match the scaling
exponents $\tilde{Z}_{p}$. The
closest fits are obtained for $\Delta=1$, where
$C_{\mathrm{sol}}\approx 1.18$ and $C_{\mathrm{comp}}\approx 1.08$,
while the models with fixed co-dimension $C=1$ yield
$\Delta_{\mathrm{sol}}=0.90$ and $\Delta_{\mathrm{comp}}\approx 0.92$.

In summary, the scaling of turbulent supersonic velocity fields
is characterized by power laws that vary substantially with the large-scale forcing. 
This finding disproves the conventional notion of universality in the supersonic regime and
bears consequences on the theory of
turbulence-regulated star formation.
On the other hand, calculating two-point statistics of the mass-weighted velocity 
$\tilde{\bm{v}}=\rho^{1/3}\bm{v}$, we found that the influence of the forcing was
considerably reduced. Based on the corresponding formulation of the
refined similarity hypothesis~(\ref{eq:ref_sim_mw}), the scaling exponents are
very well described by log-Poisson models~(\ref{eq:scl_expn}), for which the parameter $\Delta$
is close to unity. This is the expected value for the most intense dissipative structures in supersonic turbulence.



\begin{thebibliography}{19}
\expandafter\ifx\csname natexlab\endcsname\relax\def\natexlab#1{#1}\fi
\expandafter\ifx\csname bibnamefont\endcsname\relax
  \def\bibnamefont#1{#1}\fi
\expandafter\ifx\csname bibfnamefont\endcsname\relax
  \def\bibfnamefont#1{#1}\fi
\expandafter\ifx\csname citenamefont\endcsname\relax
  \def\citenamefont#1{#1}\fi
\expandafter\ifx\csname url\endcsname\relax
  \def\url#1{\texttt{#1}}\fi
\expandafter\ifx\csname urlprefix\endcsname\relax\def\urlprefix{URL }\fi
\providecommand{\bibinfo}[2]{#2}
\providecommand{\eprint}[2][]{\url{#2}}

\bibitem[{\citenamefont{Frisch}(1995)}]{Frisch}
\bibinfo{author}{\bibfnamefont{U.}~\bibnamefont{Frisch}},
  \emph{\bibinfo{title}{{Turbulence}}} (\bibinfo{publisher}{Cambridge
  University Press}, \bibinfo{year}{1995}).

\bibitem[{\citenamefont{{Kolmogorov}}(1941)}]{Kolmog41}
\bibinfo{author}{\bibfnamefont{A.}~\bibnamefont{{Kolmogorov}}},
  \bibinfo{journal}{Akademiia Nauk SSSR Doklady} \textbf{\bibinfo{volume}{30}},
  \bibinfo{pages}{301} (\bibinfo{year}{1941}).

\bibitem[{\citenamefont{{Mac Low} and {Klessen}}(2004)}]{LowKless04}
\bibinfo{author}{\bibfnamefont{M.-M.} \bibnamefont{{Mac Low}}}
  \bibnamefont{and} \bibinfo{author}{\bibfnamefont{R.~S.}
  \bibnamefont{{Klessen}}}, \bibinfo{journal}{Reviews of Modern Physics}
  \textbf{\bibinfo{volume}{76}}, \bibinfo{pages}{125} (\bibinfo{year}{2004}).

\bibitem[{\citenamefont{{Kritsuk}
  et~al.}(2007{\natexlab{a}})\citenamefont{{Kritsuk}, {Padoan}, {Wagner}, and
  {Norman}}}]{KritPad07}
\bibinfo{author}{\bibfnamefont{A.~G.} \bibnamefont{{Kritsuk}}},
  \bibinfo{author}{\bibfnamefont{P.}~\bibnamefont{{Padoan}}},
  \bibinfo{author}{\bibfnamefont{R.}~\bibnamefont{{Wagner}}}, \bibnamefont{and}
  \bibinfo{author}{\bibfnamefont{M.~L.} \bibnamefont{{Norman}}}, in
  \emph{\bibinfo{booktitle}{{AIP Conference Proceedings: "Turbulence and
  Nonlinear Processes in Astrophysical Plasmas}}}
  (\bibinfo{year}{2007}{\natexlab{a}}), vol. \bibinfo{volume}{706}.

\bibitem[{\citenamefont{{Kolmogorov}}(1962)}]{Kolmog62}
\bibinfo{author}{\bibfnamefont{A.~N.} \bibnamefont{{Kolmogorov}}},
  \bibinfo{journal}{J. Fluid Mech.} \textbf{\bibinfo{volume}{13}},
  \bibinfo{pages}{82} (\bibinfo{year}{1962}).

\bibitem[{\citenamefont{{Dubrulle}}(1994)}]{Dub94}
\bibinfo{author}{\bibfnamefont{B.}~\bibnamefont{{Dubrulle}}},
  \bibinfo{journal}{Phys. Rev. Lett.} \textbf{\bibinfo{volume}{73}},
  \bibinfo{pages}{959} (\bibinfo{year}{1994}).

\bibitem[{\citenamefont{{She} and {Waymire}}(1995)}]{SheWay95}
\bibinfo{author}{\bibfnamefont{Z.-S.} \bibnamefont{{She}}} \bibnamefont{and}
  \bibinfo{author}{\bibfnamefont{E.~C.} \bibnamefont{{Waymire}}},
  \bibinfo{journal}{Phys. Rev. Lett.} \textbf{\bibinfo{volume}{74}},
  \bibinfo{pages}{262} (\bibinfo{year}{1995}).

\bibitem[{\citenamefont{{She} and {Leveque}}(1994)}]{SheLev94}
\bibinfo{author}{\bibfnamefont{Z.-S.} \bibnamefont{{She}}} \bibnamefont{and}
  \bibinfo{author}{\bibfnamefont{E.}~\bibnamefont{{Leveque}}},
  \bibinfo{journal}{Phys. Rev. Lett.} \textbf{\bibinfo{volume}{72}},
  \bibinfo{pages}{336} (\bibinfo{year}{1994}).

\bibitem[{\citenamefont{Boldyrev}(2002)}]{Boldyrev02}
\bibinfo{author}{\bibfnamefont{S.}~\bibnamefont{Boldyrev}},
  \bibinfo{journal}{Astrophys. J.} \textbf{\bibinfo{volume}{569}},
  \bibinfo{pages}{841} (\bibinfo{year}{2002}).

\bibitem[{\citenamefont{{Benzi} et~al.}(1993)\citenamefont{{Benzi},
  {Ciliberto}, {Tripiccione}, {Baudet}, {Massaioli}, and {Succi}}}]{BenzCil93}
\bibinfo{author}{\bibfnamefont{R.}~\bibnamefont{{Benzi}}},
  \bibinfo{author}{\bibfnamefont{S.}~\bibnamefont{{Ciliberto}}},
  \bibinfo{author}{\bibfnamefont{R.}~\bibnamefont{{Tripiccione}}},
  \bibinfo{author}{\bibfnamefont{C.}~\bibnamefont{{Baudet}}},
  \bibinfo{author}{\bibfnamefont{F.}~\bibnamefont{{Massaioli}}},
  \bibnamefont{and} \bibinfo{author}{\bibfnamefont{S.}~\bibnamefont{{Succi}}},
  \bibinfo{journal}{Phys. Rev. E} \textbf{\bibinfo{volume}{48}},
  \bibinfo{pages}{29} (\bibinfo{year}{1993}).

\bibitem[{\citenamefont{{Federrath} et~al.}(2008)\citenamefont{{Federrath},
  {Klessen}, and {Schmidt}}}]{FederSchm08}
\bibinfo{author}{\bibfnamefont{C.}~\bibnamefont{{Federrath}}},
  \bibinfo{author}{\bibfnamefont{R.}~\bibnamefont{{Klessen}}},
  \bibnamefont{and} \bibinfo{author}{\bibfnamefont{W.}~\bibnamefont{{Schmidt}}}
  (\bibinfo{year}{2008}), \bibinfo{note}{in preparation}.

\bibitem[{\citenamefont{Colella and Woodward}(1984)}]{ColWood84}
\bibinfo{author}{\bibfnamefont{P.}~\bibnamefont{Colella}} \bibnamefont{and}
  \bibinfo{author}{\bibfnamefont{P.~R.} \bibnamefont{Woodward}},
  \bibinfo{journal}{J. Comp. Phys.} \textbf{\bibinfo{volume}{54}},
  \bibinfo{pages}{174} (\bibinfo{year}{1984}).

\bibitem[{\citenamefont{{Fryxell} et~al.}(2000)\citenamefont{{Fryxell},
  {Olson}, {Ricker}, {Timmes}, {Zingale}, {Lamb}, {MacNeice}, {Rosner},
  {Truran}, and {Tufo}}}]{FryOls00}
\bibinfo{author}{\bibfnamefont{B.}~\bibnamefont{{Fryxell}}},
  \bibinfo{author}{\bibfnamefont{K.}~\bibnamefont{{Olson}}},
  \bibinfo{author}{\bibfnamefont{P.}~\bibnamefont{{Ricker}}},
  \bibinfo{author}{\bibfnamefont{F.~X.} \bibnamefont{{Timmes}}},
  \bibinfo{author}{\bibfnamefont{M.}~\bibnamefont{{Zingale}}},
  \bibinfo{author}{\bibfnamefont{D.~Q.} \bibnamefont{{Lamb}}},
  \bibinfo{author}{\bibfnamefont{P.}~\bibnamefont{{MacNeice}}},
  \bibinfo{author}{\bibfnamefont{R.}~\bibnamefont{{Rosner}}},
  \bibinfo{author}{\bibfnamefont{J.~W.} \bibnamefont{{Truran}}},
  \bibnamefont{and} \bibinfo{author}{\bibfnamefont{H.}~\bibnamefont{{Tufo}}},
  \bibinfo{journal}{Astrophys. J. S.} \textbf{\bibinfo{volume}{131}},
  \bibinfo{pages}{273} (\bibinfo{year}{2000}).

\bibitem[{\citenamefont{{Eswaran} and {Pope}}(1988)}]{EswaPope88}
\bibinfo{author}{\bibfnamefont{V.}~\bibnamefont{{Eswaran}}} \bibnamefont{and}
  \bibinfo{author}{\bibfnamefont{S.~B.} \bibnamefont{{Pope}}},
  \bibinfo{journal}{Comp. Fluids.} \textbf{\bibinfo{volume}{16}},
  \bibinfo{pages}{257} (\bibinfo{year}{1988}).

\bibitem[{\citenamefont{Schmidt et~al.}(2006)\citenamefont{Schmidt,
  Hillebrandt, and Niemeyer}}]{SchmHille06}
\bibinfo{author}{\bibfnamefont{W.}~\bibnamefont{Schmidt}},
  \bibinfo{author}{\bibfnamefont{W.}~\bibnamefont{Hillebrandt}},
  \bibnamefont{and} \bibinfo{author}{\bibfnamefont{J.~C.}
  \bibnamefont{Niemeyer}}, \bibinfo{journal}{Comp. Fluids.}
  \textbf{\bibinfo{volume}{35}}, \bibinfo{pages}{353} (\bibinfo{year}{2006}).

\bibitem[{\citenamefont{{Padoan} et~al.}(2004)\citenamefont{{Padoan},
  {Jimenez}, {Nordlund}, and {Boldyrev}}}]{PadJim04}
\bibinfo{author}{\bibfnamefont{P.}~\bibnamefont{{Padoan}}},
  \bibinfo{author}{\bibfnamefont{R.}~\bibnamefont{{Jimenez}}},
  \bibinfo{author}{\bibfnamefont{A.}~\bibnamefont{{Nordlund}}},
  \bibnamefont{and}
  \bibinfo{author}{\bibfnamefont{S.}~\bibnamefont{{Boldyrev}}},
  \bibinfo{journal}{Phys. Rev. Lett.} \textbf{\bibinfo{volume}{92}},
  \bibinfo{pages}{191102} (\bibinfo{year}{2004}).

\bibitem[{\citenamefont{{Kritsuk}
  et~al.}(2007{\natexlab{b}})\citenamefont{{Kritsuk}, {Norman}, {Padoan}, and
  {Wagner}}}]{KritNor07}
\bibinfo{author}{\bibfnamefont{A.~G.} \bibnamefont{{Kritsuk}}},
  \bibinfo{author}{\bibfnamefont{M.~L.} \bibnamefont{{Norman}}},
  \bibinfo{author}{\bibfnamefont{P.}~\bibnamefont{{Padoan}}}, \bibnamefont{and}
  \bibinfo{author}{\bibfnamefont{R.}~\bibnamefont{{Wagner}}},
  \bibinfo{journal}{Astrophys. J.} \textbf{\bibinfo{volume}{665}},
  \bibinfo{pages}{416} (\bibinfo{year}{2007}{\natexlab{b}}).

\bibitem[{\citenamefont{{Alexakis} et~al.}(2005)\citenamefont{{Alexakis},
  {Mininni}, and {Pouquet}}}]{AlexaMin05}
\bibinfo{author}{\bibfnamefont{A.}~\bibnamefont{{Alexakis}}},
  \bibinfo{author}{\bibfnamefont{P.~D.} \bibnamefont{{Mininni}}},
  \bibnamefont{and}
  \bibinfo{author}{\bibfnamefont{A.}~\bibnamefont{{Pouquet}}},
  \bibinfo{journal}{Phys. Rev. Lett.} \textbf{\bibinfo{volume}{95}},
  \bibinfo{pages}{264503} (\bibinfo{year}{2005}).

\bibitem[{\citenamefont{{Fleck}}(1996)}]{Fleck96}
\bibinfo{author}{\bibfnamefont{R.~C.} \bibnamefont{{Fleck}},
  \bibfnamefont{Jr.}}, \bibinfo{journal}{\apj} \textbf{\bibinfo{volume}{458}},
  \bibinfo{pages}{739} (\bibinfo{year}{1996}).

\end{thebibliography}
\end{document}